\documentclass[prl,twocolumn,floatfix,showpacs]{revtex4}
\usepackage{graphicx}
\usepackage{amssymb,amsfonts,amsmath}
\usepackage{color}
\usepackage{ulem}

\newcommand{\nc}{\newcommand}
\nc{\rnc}{\renewcommand}
\nc{\nn}{\nonumber}

\nc{\g}{\gamma}
\nc{\om}{\omega}
\rnc{\b}{\beta}
\rnc{\th}{\theta}

\rnc{\[}{\left[}
\rnc{\]}{\right]}
\rnc{\(}{\left(}
\rnc{\)}{\right)}

\newcommand{\bra}{\langle}
\newcommand{\ket}{\rangle}
\nc{\vac}{|0\ket}
\nc{\vvac}{\bra0|}

\nc{\cd}{\cdots}
\nc{\sm}[2]{\sum_{#1=1}^{#2}}

\nc{\red}{\textcolor{red}}
\nc{\sred}[1]{\textcolor{red}{\sout{#1}}}

\nc\hp{\hat{\psi}}
\nc\hpd{\hat{\psi}^\dagger}

\nc\qs{\psi_\text{QS}}
\nc{\mfinf}{\psi_\text{MF}^{(\infty)}}
\nc{\phiinf}{\phi_\text{MF}^{(\infty)}}
\nc{\thinf}{\th_\text{MF}^{(\infty)}}
\nc{\mfperi}{\psi_\text{MF}}
\nc{\phiperi}{\phi_\text{MF}}
\nc{\thperi}{\th_\text{MF}}
\nc{\fsn}{f_{\text{sn}}}
\nc{\gsn}{g_{\text{sn}}}
\nc{\hsn}{h_{\text{sn}}}

\begin{document}
\title{
Quantum dark solitons in the 1D Bose gas and the superfluid velocity
}

\author{Jun Sato}
\author{$^\ddagger$ Rina Kanamoto}
\author{Eriko Kaminishi}
\author{Tetsuo Deguchi}
\affiliation{Department of Physics, 
Graduate School of Humanities and Sciences, 
Ochanomizu University, 2-1-1 Ohtsuka, Bunkyo-ku, Tokyo 112-8610, Japan}
\affiliation{$^\ddagger$ Department of Physics, Meiji University, 
Kawasaki, Kanagawa 214-8571, Japan}
\date{\today}

\begin{abstract}
We give explicit connections of quantum one-hole excited states 
to classical solitons 
for the one-dimensional Bose gas with repulsive short-range interactions. 
We call the quantum states connected to classical solitons 
the {\it quantum soliton states}. 
We show that the matrix element of the canonical field operator 
between quantum soliton states with $N-1$ and $N$ particles 
is given by a dark soliton of the Gross-Pitaevskii equation 
in the weak coupling case. 
We suggest that the matrix element corresponds 
to the order parameter of BEC in the quantum soliton state. 
The result should be useful in the study of many-body effects 
in Bose-Einstein condensation and superfluids. For instance, we derive the superfluid velocity 
for a quantum soliton state. 
\end{abstract} 
\maketitle
The experimental realization of trapped one-dimensional atomic gases 
has provided a new motivation in the study of the effects of 
strong correlations in fundamental quantum mechanical systems 
of interacting particles \cite{Ketterle,Esslinger,Kinoshita}. 
Furthermore, localized excitations in quantum many-body systems 
such as in cold atoms and optical lattices 
have recently attracted much interest and have been 
studied extensively in terms of ``quantum solitons'' \cite{Carr,Rina}. 
Localized quantum states are important and 
useful for investigating quite complicated excited states of 
interacting quantum systems.  
However, it is not clear how we can construct or characterize 
quantum states associated with solitons for many-body systems. 
Originally, solitons are special solutions 
of some classical nonlinear partial-differential equations. 
It is not even trivial to see whether there exists 
a quantum state with a soliton-like density profile.

Let us consider the Gross-Pitaevskii (GP) equation which describes 
Bose-Einstein condensation (BEC) 
in the mean-field approximation \cite{Leggett}. We also call it 
the classical nonlinear Schr{\"o}dinger equation, 
since it corresponds to the classical limit of the quantum nonlinear Schr{\"o}dinger equation 
satisfied by the canonical Bose field ${\hat \psi}(x,t)$ for 
the one-dimensional Bose gas interacting with the delta-function potentials. 
Here, the system is called the Lieb-Liniger (LL) model \cite{Lieb-Liniger}. 
The GP equation has dark soliton solutions 
for the repulsive interactions, while it has 
bright soliton solutions for the attractive interactions \cite{soliton}.  
It was conjectured that dark solitons are identified with  
Lieb's type-II excitations as an excitation branch \cite{Takayama}. 
However, it has not been shown 
how one can construct such quantum states 
that are related to solitons or 
what kind of physical quantity can 
show a property of solitons for some states. 
In fact, each of the type-II eigenstates has a flat density profile, 
since the Bethe eigenstates are translationally invariant. 
Here we remark that for the attractive case, 
bright solitons are analytically derived 
from some quantum states of the LL model \cite{Wadati}. 

In this Letter we demonstrate that quantum states which are 
tightly connected to classical solitons 
are constructed from the Bethe eigenvectors 
of the LL model. We call the states the {\it quantum soliton states}. 
Let us denote by $\qs(x)$ 
the matrix element of the field operator ${\hat \psi}(x, t)$ 
between two quantum soliton states where 
one state has $N-1$ particles and another $N$ particles. 
We show that the matrix element $\qs(x)$ 
is well approximated by the classical complex scalar field 
of a dark soliton of the GP equation in the weak coupling case. 
We suggest that the matrix element $\qs(x)$ 
corresponds to the order parameter of BEC 
in the system with a large but finite number of interacting particles 
in the weak coupling case. 
The result should be fundamental in the study of many-body effects 
in BEC and superfluids. For an illustration, we derive the superfluid velocity 
from the phase profile of the matrix element  $\qs(x)$.

We give remarks. 
First, superposing Lieb's type II excitations, i.e. one-hole excitations,  
we construct the quantum soliton states \cite{SKKD1}, 
which have broken translational symmetry. 
The Bethe eigenstates are translationally invariant, 
while their superpositions are not, in general.  
Secondly, we show that the amplitude and phase profiles of 
quantum soliton states are consistent with those 
of corresponding solitons of the GP equation. 
Although the density profile with a density notch  
has been derived for a quantum soliton state \cite{SKKD1}, 
the connection to solitons has not been shown, yet.   
Thirdly, it is not {\it a priori} clear 
how valid  the mean-field approximation is   
for the quantum soliton states. 
The exact wavefunctions given by the Bethe ansatz 
consist of a large number of terms such as $N!$.  
However, evaluating the matrix element $\qs(x)$ we identify it 
as the order parameter of BEC.

Let us consider the Hamiltonian of the LL model \cite{Lieb-Liniger}: 
\begin{align}
{\cal H}_{\text{LL}} 
= - \sum_{j=1}^{N} {\frac {\partial^2} {\partial x_j^2}}
+ 2c \sum_{j < k}^{N} \delta(x_j-x_k) . 
\end{align}
Here the periodic boundary conditions (P.B.C.) 
of the system size $L$ are assumed on the wavefunctions. 
Hereafter, we consider the repulsive interaction: $ c > 0 $. The LL model is 
characterized by a single parameter $\g:=c/n$, 
where $n=N/L$ is the particle density. 
We employ a system of units with $2m=\hbar =1$, 
where $m$ is the particle mass. 
The second-quantized Hamiltonian of the LL model 
is written in terms of the canonical Bose field $\hat{ \psi}(x,t)$ as
\begin{align}
{\cal H}_{\text{NLS}} = 
\int_{0}^{L} dx [ \partial_x \hat{ \psi}^{\dagger} \partial_x \hat{ \psi} + 
c \hat{ \psi}^{\dagger} \hat{ \psi}^{\dagger} \hat{ \psi} \hat{ \psi} - 
\mu \hat{ \psi}^{\dagger} \hat{ \psi} ],  
\end{align}
where $\mu$ is the chemical potential. The Heisenberg equation of motion 
is called the nonlinear Schr{\"o}dinger equation: 
$i \partial_t \hat{ \psi} =  - \partial^2_x \hat{ \psi} 
+ 2c \hat{ \psi}^{\dagger} \hat{ \psi} \hat{ \psi} 
- \mu \hat{ \psi}. $ 

In the LL model, the Bethe ansatz offers an exact eigenstate 
with an exact energy eigenvalue for a given set of quasi-momenta 
$k_1, k_2, \ldots, k_N$ satisfying 
the Bethe equations 
for $j=1, 2, \ldots, N$: 
\begin{align}
k_j L = 2 \pi I_j - 2 \sum_{\ell \ne j}^{N} 
\arctan \left({\frac {k_j - k_{\ell}} c } \right). 
\label{BAE} 
\end{align}
Here $I_j$'s are integers for odd $N$ and half-odd integers for even $N$. 
We call them the Bethe quantum numbers. 
The total momentum $P$ and energy eigenvalue $E$ are written 
in terms of the quasi-momenta as 
$ P=\sm{j}{N}k_j=\frac {2 \pi} L \sum_{j=1}^{N} I_j, \quad E=\sm{j}{N}k_j^2. $
If we specify a set of Bethe quantum numbers 
$I_1<\cdots<I_N$, the Bethe equations 
\eqref{BAE} have 
a unique real solution $k_1 < \cdots < k_N$ \cite{Korepin}. 

Let us formulate quantum soliton states \cite{SKKD1}. 
We shall show throughout the Letter 
that they lead to dark solitons of the GP equation.  
In the type II branch, for each integer $p$ in the set $\{0, 1, \ldots, N-1\}$, 
we consider momentum $P=2 \pi p/L$
and denote by $|P, N \ket$ the normalized Bethe eigenstate of $N$ particles 
with total momentum $P$. 
The Bethe quantum numbers of $|P, N \ket$ are given by
$I_j=-(N+1)/2+j$ for integers $j$ with $1\leq j \leq N-p$ and 
$I_j=-(N+1)/2+j+1$ for $j$ with $N-p+1\leq j \leq N$. 
For each integer $q$ satisfying $0 \le q \le N-1$ 
we define the coordinate state $|X, N \ket$ of $X=qL/N$
by the discrete Fourier transformation: 
\begin{align}
| X, N \ket := \frac 1 {\sqrt{N}} \sum_{p=0}^{N-1} \exp(- 2 \pi i p q/N) \, | P, N \ket \, . 
\label{eq:XN}
\end{align}
\begin{figure}[t]
\includegraphics[width=0.8\columnwidth]{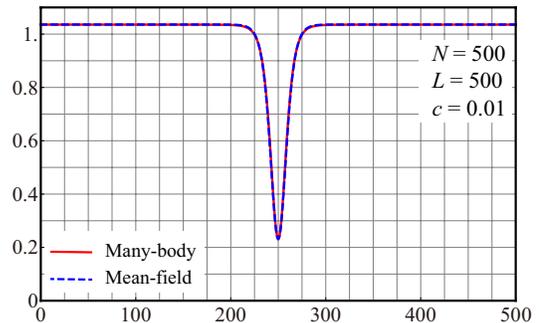}
\caption{(Color online) 
Density profile of the quantum soliton 
state $\bra X, N|\hpd(x)\hp(x)|X, N\ket$ 
for $c=0.01$ and $N=L=500$ 
is shown  by a red solid line. 
The profile of the squared amplitude of a dark soliton $|\mfperi(x)|^2$ 
with $v \simeq v_c/2$ is plotted with a blue broken line. 
}
\label{density}
\end{figure}

The density profile of the quantum soliton state,  
$\bra X, N|\hpd(x)\hp(x)|X, N\ket$ versus $x$, 
is  plotted in Fig. \ref{density}. 
It is denoted by ``Many-body''.
Here we have set the coordinate integer as $q=0$, 
and  the density notch is localized at $x=L/2$. 
The expectation values of the density operator 
are effectively calculated \cite{SKKD1} by
the determinant formula for the norms of Bethe eigenstates \cite{GK} 
and that of the form factors of the density operator 
\cite{Slavnov, Caux2007}.

The classical complex scalar field  of a dark soliton solution 
for the GP equation with P.B.C., $\mfperi(x)$ ($0 \le x \le L$), 
is derived  by 
assuming the traveling-wave solution: $\psi(x, t)= \mfperi(x-vt)$.  
Here we note that the periodic soliton solutions of the GP equation 
are expressed in terms of the elliptic integrals \cite{Rina}. 
We also note that the excitation mode has the largest velocity $v_c$,  
which we  call the critical velocity: there is no soliton solution with 
$v >v_c$ \cite{Rina,soliton,Takayama}. 
In the LL model,  the critical momentum $p_c=m v_c$ 
corresponds to the Fermi momentum $2 k_F$.

The density profile of the quantum soliton state and 
the square-amplitude profile of a classical dark soliton  
with P.B.C., $|\mfperi(x)|^2$,  
agree quite well in the weak coupling case $c \ll 1$, 
as shown in Fig. \ref{density}. 
Here we have set $v \simeq v_c/2$ for the dark soliton, and 
its profile is denoted by ``Mean-field''.

We suggest that the soliton velocity $v \simeq v_c/2$ is consistent with 
the construction (\ref{eq:XN}) of the quantum soliton state. 
Each of the type II excitations 
in the range $0\leq v \leq v_c$ is superposed with equal weight, 
so that we have the average value $v \simeq v_c/2$. 
It looks like a wave packet of the type II excitations. 

Let us consider the matrix element of the field operator 
between the quantum soliton states   
\begin{align}
&\qs(x):=
\bra X, N-1|\hp(x)|X, N \ket
\nn\\&
=
\frac1{\sqrt{N(N-1)}}
\sum^{N-1}_{p=0}
\sum^{N-2}_{p'=0}
e^{i(P-P')x}
\bra P', N-1|\hp(0)|P, N\ket, \label{eq:matrix}
\end{align}
where $P=2\pi p/L$ and $P'=2\pi p'/L$ denote the total momenta 
of the normalized Bethe eigenstates 
$|P, N\ket$ and $|P', N\ket$, respectively. 
We put $q=0$ in eq. (\ref{eq:matrix}). 
The matrix element $\bra P', N-1|\hp(0)|P, N\ket$ 
are evaluated effectively by the determinant formula for the 
norms of Bethe eigenstates \cite{GK} and that 
for the form factors of the field operator 
\cite{Slavnov, Kojima, Caux-Calabrese-Slavnov2007} as
\begin{align}
&\bra P', N-1|\hp(0)|P, N\ket
=(-1)^{N(N+1)/2+1}
\nn\\&\times
\(
\prod^{N-1}_{j=1}
\prod^N_{\ell=1}
\frac{1}{k'_j-k_\ell}\) 
\( 
\prod^N_{j>\ell}k_{j,\ell}\sqrt{k_{j,\ell}^2+c^2} 
\)
\nn\\&\times
\( 
\prod^{N-1}_{j>\ell}\frac{k'_{j,\ell}}{\sqrt{(k'_{j,\ell})^2+c^2}} 
\)
\frac{\det U(k,k')}{\sqrt{\det G(k)\det G(k')}}, 
\label{eq:Slavnov}
\end{align}
where the quasi-momenta $\{k_1,\cdots,k_N\}$ and $\{k'_1,\cdots,k'_{N-1}\}$ 
give the eigenstates $|P, N\ket$ and $|P', N-1\ket$, respectively. 
Here we have employed the abbreviated symbols 
$k_{j,\ell}:=k_j-k_\ell$ and $k'_{j,\ell}:=k'_j-k'_\ell$. 
The matrix $G(k)$ is the Gaudin matrix, whose $(j,\ell)$th element is 
$G(k)_{j,\ell}=\delta_{j,\ell}\[L+\sum_{m=1}^NK(k_{j,m})\]-K(k_{j,\ell})$
for $j, \ell=1,2,\cdots,N$, where the kernel $K(k)$ 
is defined by $K(k)=2c/(k^2+c^2)$. 
The matrix elements of the $(N-1)$ by $(N-1)$ matrix $U(k,k')$ are given by 
\begin{align}
U(k,k')_{j,\ell}&=2\delta_{j\ell}\text{Im}\[
\frac{\prod^{N-1}_{a=1}(k'_a-k_j + ic)}{\prod^N_{a=1}(k_a-k_j + ic)}\]
 \nn\\&
+\frac{\prod^{N-1}_{a=1}(k'_a-k_j)}{\prod^N_{a\neq j}(k_a-k_j)}
\(K(k_{j,\ell})-K(k_{N,\ell})\). 
\label{eq:matrixU}
\end{align}

\begin{figure}[t]
\includegraphics[width=0.99\columnwidth]{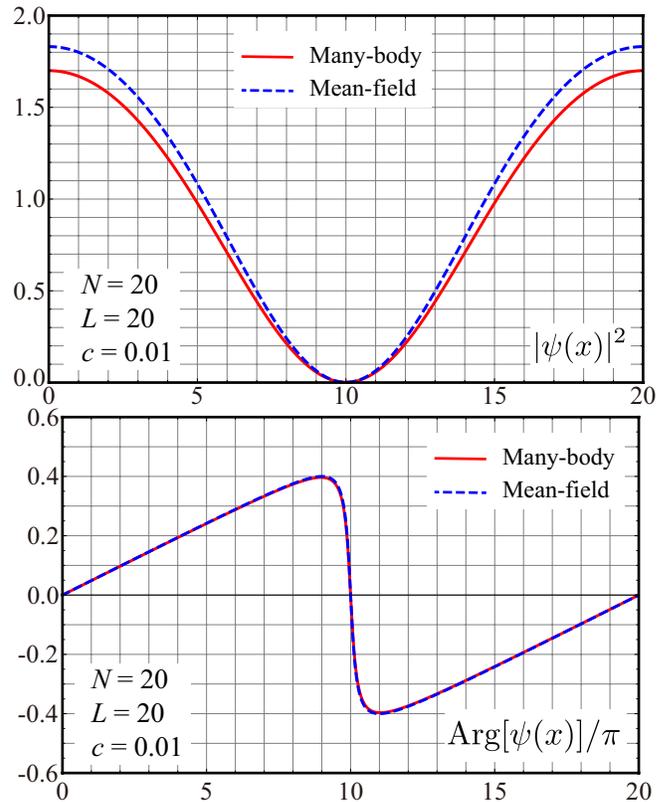}
\caption{(Color online) 
Profiles of the squared amplitude $|\qs(x)|^2$ and 
the phase $\text{Arg}[\qs(x)]/\pi$ 
for the matrix element of the field operator, $\qs(x)$,  
are shown by red solid lines 
for $c=0.01$ and $N=L=20$. Those of a dark soliton under P.B.C., $\mfperi(x)$, 
with $v \simeq 2\pi/L$ are plotted with blue dotted lines. Here we put $q=0$. }
\label{field20}
\end{figure}

\begin{figure}[t]
\includegraphics[width=0.99\columnwidth]{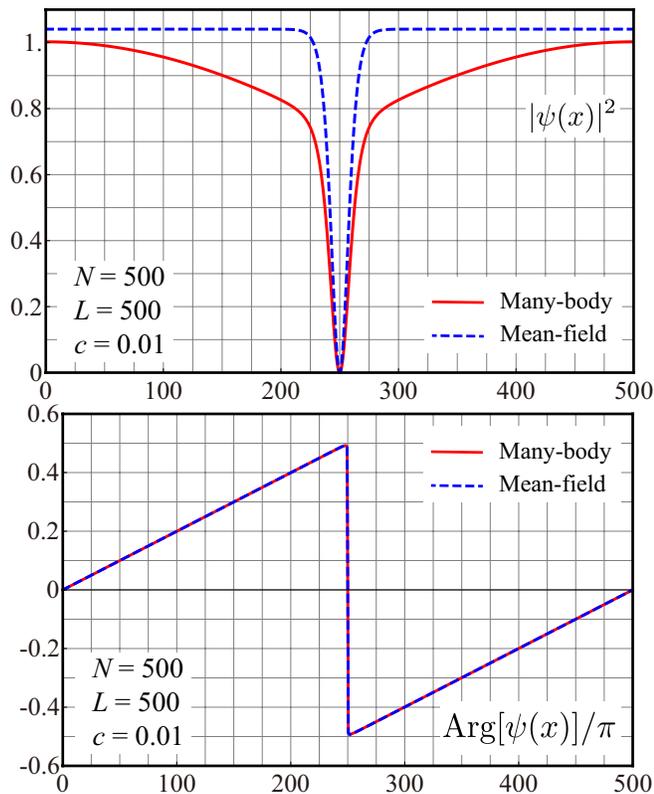}
\caption{(Color online) 
Same plots as in Fig. \ref{field20} with the system size $N=L=500$. 
}
\label{field500}
\end{figure}

The profiles of the squared amplitude $|\qs(x)|^2$ 
and the phase  $\text{Arg}[\qs(x)]/\pi$ 
are plotted in Figs. \ref{field20} and \ref{field500} for 
$N=20$ and 500, respectively.   
The squared amplitude and phase profiles of 
periodic dark solitons, $\mfperi(x)$,  
are shown by broken blue lines in Figs. \ref{field20} and \ref{field500}  
for $N=20$ and 500, respectively. 
They have the velocity $v \simeq 2 \pi/L$. 
Here the classical complex scalar field $\mfperi(x)$ is normalized such that  
the integral of $|\mfperi(x)|^2$ with $x$ 
over the whole region gives the particle number $N$.

The matrix element of the field operator, 
$\qs(x)$, and the classical dark soliton with P.B.C.,   
$\mfperi(x)$, are in good agreement around 
at the central part of the solitons 
in Figs. 2 and 3. In particular, 
the phase profiles of $\qs(x)$ and $\mfperi(x)$ 
completely overlap each other (see, the lower panels of Figs. 2 and 3).  
The profiles of square-amplitude $|\qs(x)|^2$ of the quantum soliton states 
are slightly smaller than those of the periodic solitons, 
$|\mfperi(x)|^2$, (see, the upper panels of Figs. 2 and 3).       
For $N=20$ and $c=0.01$ the two profiles are proportional to each other    
only with different normalizations. For $N=500$ and $c=0.01$ the two profiles 
overlap each other at the central part and deviate around at the shoulders.

The agreement of the squared amplitudes 
$|\qs(x)|^2$ and $|\mfperi(x)|^2$ 
should be improved for smaller values of $c$. 
Let us consider the form-factor expansion 
of the local density at $x$ for the state $|X, N\ket$:  
\begin{eqnarray} 
& &  \bra X, N| \hpd(x) \hp(x)|X, N \ket 
\nonumber \\ 
&  & = 
|\qs(x)|^2 +  
\sum_{|n \ket  \ne |X, N-1 \ket} 
|\bra n| \hp(x)|X, N\ket|^2.  \label{eq:decomp}
\end{eqnarray} 
The second terms of the right hand side of (\ref{eq:decomp}) 
give corrections to $|\qs(x)|^2$, and    
the integral of $|\qs(x)|^2$ with $x$ 
over the whole region is smaller than the particle number $N$. 
We observed numerically that the correction terms become small 
as the coupling constant $c$ decreases if we fix the particle number $N$, 
while they increase as $N$ increases if  $c$ is fixed. 
In the large $N$ case, the correction terms should be small 
if the value of $c$ is small enough. 
We thus conclude that the matrix element $\qs(x)$ 
is well approximated by the periodic dark soliton $\mfperi(x)$ 
in the weak coupling case.

We suggest that the soliton velocity $v \simeq 2 \pi/L$  
corresponds to the difference between 
the average values of the total momenta of the quantum soliton states 
$|X, N \ket$ and $|X, N-1 \ket$.  It is 
consistent with the structure of the matrix element  
$\bra X, N-1|\hp(x)|X, N \ket$.

%

We now argue for the claim that the matrix element $\qs(x)$ 
gives the order parameter of BEC 
in the quantum soliton state $|X, N \ket$ 
for the weak coupling and large-$N$ case. 
Here the system size $L$ is also very large since we set $n=N/L=1$.  
We denote by $\rho_1(x, y)_{| \Psi \ket}$  
the one-particle reduced density matrix for a given state $| \Psi \ket$: 
$\rho_1(x, y)_{| \Psi \ket} = \bra \Psi| \hpd(x) \hp(y) | \Psi \ket$.   
We now conjecture that the matrix element $\qs(x)$ satisfies the relation: 
${\rho}_1(x, y)_{|X, N \ket} \simeq \qs^{*}(x) \qs(y)$ for $|x-y| \gg 1$. 
We call it conjecture A. 
Here we  assume that the system size $L$ 
is much larger than the healing length $l_c= 1/ \sqrt{cn}$. 
For instance, $\ell_c=10$ for $c=0.01$ and $n=1$.

Suppose that the one-particle reduced density matrix 
$\rho_1(x,y)_{|\Psi \ket}$ for a given state $|\Psi \ket$ is diagonalized as   
\begin{equation}
\rho_1(x,y)_{|\Psi \ket} 
= \sum_{i} n_i \, \chi^{*}_{i}(x) \, \chi_{i}(y) . \label{eq:diag}
\end{equation}
For the ground state we can numerically show 
that for small $c$ the largest eigenvalue $n_0$ 
of $\rho_1(x,y)$ is much larger than 
the other eigenvalues: $n_0 \gg n_i$ for $i \ne 0$,  
i.e. the existence of BEC \cite{SKKD1}.  
For the state $|X, N \ket$ it could be technically nontrivial
to diagonalize $\rho_1(x,y)_{|X, N \ket}$ numerically. 
However, instead of doing it we point out that 
conjecture A is consistent with the following observation:  
The quantum soliton state $|X, N-1 \ket$ is dominant 
among the intermediate states in the expansion (\ref{eq:decomp}). 
Here, we estimate the fraction of the correction term  
from the difference between the local density at $x$ 
for the state $|X, N\ket$ and the squared amplitude $|\qs(x)|^2$, 
and it is small for small $c$ and large $N$. 
Moreover, from the difference we estimate   
the condensate depletion, i.e. the fraction of the non-condensate components. 
It should have the largest values for $x=y$, since 
the local density at $x$ gives the diagonal element of 
$\rho_1(x, y)_{|X, N \ket}$ with $x=y$.

We therefore conjecture that for small $c$ and large $N$ 
the order parameter $\sqrt{n_0} \chi_0(x)$ 
of the quantum soliton state $|X, N\ket$ 
is given by the matrix element $\qs(x)$, 
which is well approximated by 
the periodic dark soliton $\mfperi(x)$ with $v \simeq 2\pi/L$.   
Here, the order parameter $\sqrt{n_0} \chi_0(x)$ has been 
defined by eq. (\ref{eq:diag}) for $|\Psi \ket = |X, N\ket$. 
In terms of BEC 
we have connected the dark soliton $\mfperi(x)$ with 
$v \simeq 2 \pi/L$ to the state  $|X, N\ket$. 
It was not trivial to specify the soliton velocity $v$.

Let us now derive the superfluid velocity 
for a quantum soliton state $| X, N \ket$. 
For large $N$ such as $N = 500$, 
the phase field is fitted by $\theta(x)=\pi x/L - \pi H(x-X-L/2)$,  
as shown in the lower panel of Fig. 3. 
Here $H(x)$ denotes Heaviside's step function: 
$H(x)= 1$ for $x\ge 0$, and $H(x)= 0$ otherwise.  
Numerically we observed that the phase profile does not 
depend on the value of $c$ for large $N$ and small $c$. 
We derive the superfluid velocity 
from the phase field of the macroscopic wavefunction, $\theta(x)$, 
by $v_s = 2 ({\hbar}/{2m}) ({d \theta}/{d x})$ \cite{QL}. 
For large $N$ we thus have 
\begin{equation} 
v_{s}= {\frac {2 \pi} L} - \pi \,  \delta(x-X-L/2) . 
\end{equation}
The superfluid velocity $v_{s}$ 
has a singularity at the location of 
the soliton, and is consistent with the 
soliton velocity $v \simeq 2 \pi/L$.

The finding in the Letter suggests several possible future researches 
in quantum dynamics such as the collision of two quantum solitons, 
which is nontrivial in the dynamics 
of interacting quantum systems \cite{Drummond}.

In conclusion, in order to prove that 
the quantum states $|X, N \rangle$ constructed in (\ref{eq:XN}) 
for the 1D Bose gas are closely connected to classical solitons,   
we have shown the following two points:   
First, the density profile of the state $|X, N \rangle$ is consistent with 
the profile of the squared amplitude $|\mfperi(x)|^2$ 
of the periodic dark soliton of the GP equation with $v \simeq v_c/2$. 
Then, the matrix element of the Bose field operator, 
$\qs(x)= \bra X, N-1 | \hp(x) |X, N \ket$,  
 coincides with the classical complex scalar field $\mfperi(x)$ 
of the dark soliton of the GP equation 
under P.B.C. with $v \simeq 2 \pi/L$.  
The agreement is good for small $c$. 
Furthermore, we suggest that the matrix element $\qs(x)$ 
gives the orer parameter of BEC 
in the quantum soliton state 
$|X, N \rangle$ for small $c$ and large $N$.

The authors thank I. Danshita and K. Sakai 
for their useful discussions. 
J.S. and E.K. are supported by JSPS.

\end{document}